\documentclass[secnumarabic, prc, showkeys, showpacs]{revtex4}

\usepackage{bm}
\usepackage{amsmath,amsfonts}
\usepackage[english]{babel}
\usepackage{graphicx}
\begin{document}

\title{Physical interpretation of the fringe shift measured on Michelson interferometer in optical media}
\author{V.V.Demjanov}
\affiliation{Ushakov State Maritime Academy, Novorossyisk, Russia}
\email{demjanov@nsma.ru}
\date{\today}

\begin{abstract}
 The shift $\Delta X_m$ of the interference fringe in the Michelson interferometer is absent when arm's light carriers are  vacuum ($n=1$) but present in measurements where the refractive index $n$ of its light carrying media $n> 1$. This experimental observation induced me to interpret physical processes occurred in the Michelson interferometer in a conceptually new way.  I rejected the classical rule $c'=c+v$ of adding velocity $v$ on an inertial source of the interferometer with the velocity $c$ of the light emitted by it as inapplicable in principle to non-inertial object which electromagnetic waves just belong to, and not taking into account $n$ of the optical medium. I used instead of this rule the non-relativistic formula of Fresnel $c/n \pm v(1-1/n^2)$ for the speed of light in a moving optical medium, where $n$ is taken into account and the condition $c'<c$ always holds.  This formula, and accounting for the physical effect of Lorentz contraction of the longitudinal arm $l_\parallel$ of interferometer because of its motion in the stationary aether, enabled me to construct the theoretical model $\Delta X_m \sim l\Delta\varepsilon(1-\Delta\varepsilon)$, where $\Delta\varepsilon = n^2-1$ is the contribution of particles into the permittivity of light carriers,  that reproduced  in essential features the parabolic dependence of $\Delta X_m$ on $\Delta\varepsilon$. From the experimentally measured amplitudes of shifts of the interference fringe there was estimated the horizontal projection of the velocity of the Earth relative to aether that changes in different times of day and night at the latitude of Obninsk within the limits $140\div480$ km/s.

 The forth version of the report is complemented with the account of the harmful effect on registering the shift $\Delta X_m$ of the interference fringe of the improperly installed in turning-points of the zigzag light path glass mirrors. The nullifying of the shift in air interferometers may occur because of opposite signs of contributions of the air and glass optical media through the discovered by me law $\Delta X_m \sim l\Delta\varepsilon(1-\Delta\varepsilon)$. This may be the cause of obtaining negative results in majority of Michelson-type experiments known to date.
\end{abstract}
\maketitle

\section{Michelson experiment and its standard interpretation}

By definition, aether is a hypothetical medium that serves to carry over electromagnetic waves and to transmit interactions. Supposedly the Earth moves through the luminiferous aether with a velocity ${\bf v}$.  In order to detect this motion experimentally Michelson \cite{Michelson, Michelson Morley} measured the time $t$ needed for the light to cover the distance $l$ from the light's source to the rebounding mirror forth and back in two directions: $t_\|$ $-$ in the arm parallel to ${\bf v}$ and $t_\perp$ $-$ in the arm perpendicular to ${\bf v}$. The experimental data obtained were interpreted by him in the following way. He considered it will be sufficient to determine the time of the light's propagation in the longitudinal arm of the length $l_\|$ forth by the velocity $(c+v)$, and back by the velocity $(c-v)$ in the forms not taking into account the refractive index of the optical medium \cite{Michelson Morley}:
\begin{equation}
t_\|=\frac{l_\|}{c-v}+\frac{l_\|}{c+v}=\frac{2l_\|}{c}\frac{1}{1-v^2/c^2}\approx\frac{2l_\|}{c}(1+\frac{v^2}{c^2}).\label{longitudinal}\\
\end{equation}
The time that the light spreads in the transverse arm, accounting for the Lorentz correction for the motion of the beam along the hypotenuses of the triangle, is usually written as:
\begin{equation}
t_\perp=\frac{2l}{\sqrt{c^2-v^2}}=\frac{2l}{c}\frac{1}{\sqrt{1-v^2/c^2}}\approx\frac{2l}{c}(1+\frac{1}{2}\frac{v^2}{c^2}).\label{transverse}
\end{equation}
For the case in question measuring the difference of times (\ref{transverse}) and (\ref{longitudinal}), e.g. for $l_\|=l_\perp =l$,
\begin{equation}
\Delta t=t_\perp -t_\|\approx-\frac{v^2}{c^2}\frac{l}{c}\label{Michelson}
\end{equation}
we would find the speed $v$ of the "aether wind". Michelson used Maxwell's idea to determine $\Delta t$ by  measuring the amplitude of shift $\Delta X_m$ of the interference fringe in the superposition of two beams at the interference screen. Amplitude of the shift  $\Delta X_m$ is related with  $\Delta t$ by the proportion \cite{Demjanov}:
\begin{equation}
\Delta X_m=cX_o\Delta t/\lambda\label{shift}
\end{equation}
where $X_o$ is the width of the interference fringe, and $\lambda$ the wavelength of the light's ray of the interferometer.

However the measurements of Michelson and later experiments showed that the value of $\Delta X_m$ "equals to null". This fact has been explained in the bounds of the physical effect of Lorentz-Fitzgerald contraction (longitudinal to $\mathbf{v}$) of the length $l_\|$ of a moving with the velocity $v$ body:
\begin{equation}
l_\|=l\sqrt{1-v^2/c^2}\label{contraction}
\end{equation}
where $l_\|$ is the length of the longitudinal arm of the moving interferometer. In this model the length $l_\perp$ was taken to be independent on $\mathbf{v}$. With the account of (\ref{contraction}) we obtain instead of  (\ref{longitudinal})
\begin{equation}
t_\|=\sqrt{1-v^2/c^2}\cdot\frac{2l}{c}\frac{1}{1-v^2/c^2}=\frac{2l}{c}\frac{1}{\sqrt{1-v^2/c^2}}.\label{longitudinal contraction}
\end{equation}
From (\ref{longitudinal contraction}) and (\ref{transverse}) we have the exact equality $t_\|=t_\perp$, i.e. $\Delta t = 0$.

\section{Michelson experiment in optical media and its new  relativistic interpretation}

In 1968-1974 being a scientific researcher at the Obninsk branch of the Karpov Institute of Physical Chemistry and experimenting with the Michelson interferometer, I detected the occurrence of the nonvanishing amplitude of the shift ($\Delta X_m\neq 0$) of the interference fringe in the air of normal and enhanced pressure that continually vanished in the course of the evacuation of the air from light carrying regions of the interferometer \cite{Demjanov}. I measured $\Delta t$ for various transparent dielectric materials with the optical dielectric permittivity $0<\varepsilon<3.5$ that I have managed recently to briefly publish \cite{Demjanov}-\cite{Demjanov first order}.

\begin{figure}[h]
  \begin{center}
 \includegraphics[scale=0.7]{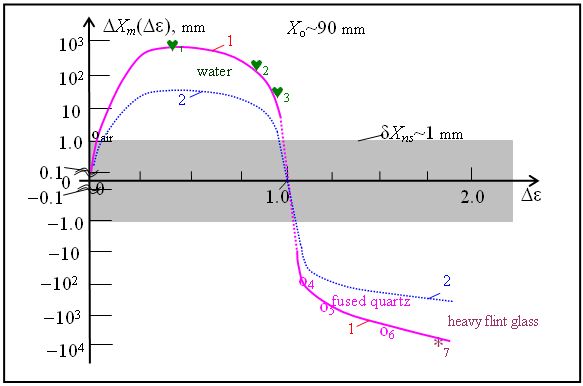}
  \caption{Dependence of amplitude of the harmonic component of the interference fringe shift $\Delta X_m$ on the contribution $\Delta\varepsilon$ of the dielectric permittivity of particles to full dielectric permittivity $\varepsilon=1+\Delta\varepsilon$ of optical medium: air ($\circ_{air})$, water (solid $\heartsuit$), fused quartz ($\circ$), heavy flint glass ($*$). Curve 1 corresponds to maxima of $\Delta X_m$, and curve 2 to minima of $\Delta X_m$ over the period of 24-hour observation. At the Obninsk latitude the curve 1 is observed during $1\div1.5$ hours in June at 12 o'clock, in September at 18 o'clock, in December at 24 o'clock, in March at 6 o'clock Moscow time. The curve 2 represents amplitudes $\Delta X_m$ about 14 times lesser than on curve 1 since the values $X_m=X_{m min}$ for the curve 2 is observed with the shift in 12 hours (Moscow time) relative to times of observation of respective points $X_m=X_{m max}$ for the curve 1. Measurements of $\Delta X_m(\Delta\varepsilon)$  for eighth values of $\Delta\varepsilon$ were performed in 1968$-$1971 years on interferometers with different lengths $l$ of the light carrying medium: for gases the length of light carriers was $l=l_\|=l_\perp=6$ m, for water and solids the length of light carriers was $l=l_\|=l_\perp=0.3$ and 0.1 m. The point $\circ_{air}$ was measured at the wavelength of the light's source $\lambda=6\cdot10^{-7}$ m, and points 1, 4; 2, 5 and 3, 6, 7 were obtained at $\lambda\sim7\cdot10^{-6}, 9\cdot10^{-7}$ and $\lambda\sim3\cdot10^{-6}$ m, respectively.  $X_0$ is the interference bandwidth, $\delta X_{ns}\sim1$ mm is the level of the  noise-jitter of the interference fringe. All data in the Figure are reduced to $l_0=6$ m and $\lambda_0=6\cdot10^{-7}$ m.}\label{fig1}
\end{center}
\end{figure}

Fig.\ref{fig1} presents more detailed experimental data comparing with those that I have given in e-prints arXiv: 0910.5658, v1 and v2 (see \cite{Demjanov second order}). The necessity to refine them (re-normalization to common parameters: length  $l$ of arms, wavelength  $\lambda$ and a single time of day or night of implementing the measurements) was pointed out to me by referees of the manuscript of the article from the journal Phys.Lett.A \cite{Demjanov PLA}. I thoroughly performed such re-normalization in Fig.1 of  \cite{Demjanov PLA} (after which the article was accepted for publication in PLA). And so I presents below this figure from PLA in a more elaborated form as  Fig.\ref{fig1} (supplementing it by the curve 2) with a more detailed description of it.

Because of the brevity of the previous description of my experimental results, I presents below the table 1 containing full primary (non-normalized but directly read off the screen of the kinescope) values of measured amplitude of harmonic shifts $\Delta X_m$ of the interference fringe. These data were obtained at three different frequencies of the rays, that provided me in the region of normal dispersion with six different values of the dielectric permittivity of the light carriers: three at water (points 1, 2, 3) and three at fused quartz  (points 4, 5, 6). The seventh point ($*_7$) I obtained at the glass "heavy flint" $-$ the most high permeable in my experiments. In comparison with Fig.1 of the article in \cite{Demjanov PLA} the Fig.\ref{fig1} given below is supplemented by the curve 2 corresponding to minimal values of the amplitude of the shift of the interferometer fringe in the 24-hour cycle of the day and night of the observing the value $\Delta X_{m min}$  obtained at above mentioned seven permittivities of light carrying media of my interferometer.

\begin{table}
  \centering
  \includegraphics[scale=0.55]{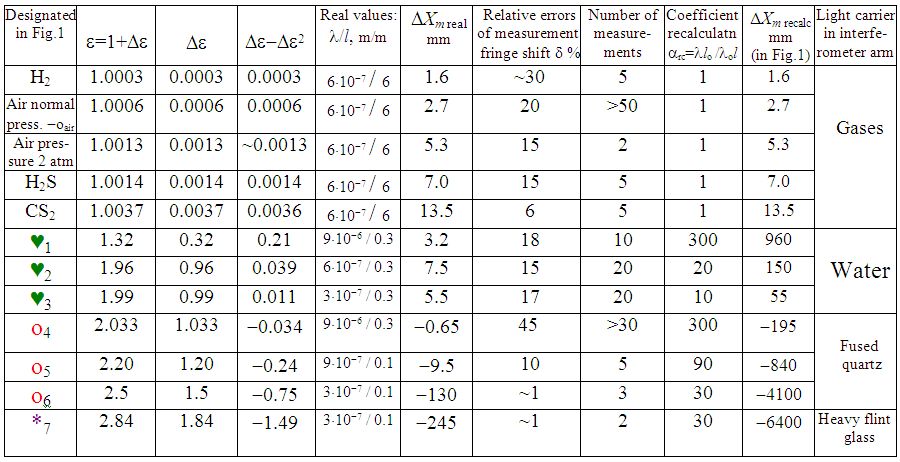}
  \caption{Actual results (in $1\div8$ columns) of 12-th series of my measurements at the Michelson-type interferometer  with various light carriers, selected by me to Fig.\ref{fig1} of the report \cite{Demjanov PLA}.}\label{table1}
\end{table}

The first my discovery has been that the pumping off the air from the region of light carrying space of the both arms of the interferometer (i.e. evacuating the light carrying space) turned entirely to null the amplitude of the harmonic shift of the interference shift (making $\Delta X_m=0$). The restoration of the normal pressure of the air in regions of the light carrying space recovered the non-zeroth value of the amplitude of the harmonic shift of the interference shift (making $\Delta X_m>0$), in which connection going on the increase of the air pressure in light carrying envelopes brought to further growth of $\Delta X_m$ (for instance, at the pressure 2 atm the value $\Delta X_m$ redoubled in comparison with the case of normal air pressure, see table 1).

In table 1 I present also the data of measuring $\Delta X_m(\varepsilon$ performed for four gaseous light carriers  covering the interval of permittivities $1.0003\leq\varepsilon\leq1.0037$. It is impossible to show these data at the linear scale of the axis $\Delta\varepsilon$ Fig.1, and so there is shown in it conditionally only the point $\circ_{air}$ for the air of normal pressure in order that the reader can see the low quality of all experimental measurements on the Michelson interferometer in the air environs which were obtained before me (up to 1968 year), since all of them were situated lower than my point $\circ_{air}$ and were drowned in the noise.  Full data of the table 1 in the form of plots in double logarithmic scale of axes $\Delta X_m$ and $\Delta\varepsilon$ were published by me in \cite{Demjanov experiment}.

In Fig.\ref{fig1} there is shown the found by me parabolic dependence of the amplitude of the fringe shift $\Delta X_m(\Delta\varepsilon)$ on the part $\Delta\varepsilon=\varepsilon-1$ of the full dielectric permittivity $\varepsilon$ of the optical medium excluding the contribution of the polarization of "vacuum"=1. As we can see from Fig.\ref{fig1} the output ratio signal/noise is improved considerably for media with higher refractive index ($n>1$ where $n=\sqrt{\varepsilon}$) than that of the air.

Observing in my experiments the dependence of the interference fringe shift on the dielectric permittivity of light carriers of Michelson interferometer, I have come in the end to the above described analysis of dispersion properties of the light carrying medium. In order to explain the run of the experimental curve  $X_m(\Delta\varepsilon)$ shown in Fig.\ref{fig1} I rejected as inadequate the rule of composition $c\pm v$ of the light's speed $c$ with the velocity $v$ of the translatory motion of the optical medium that was used by Michelson in deriving relations (\ref{longitudinal}-\ref{Michelson}). Although the Maxwell-Sellmeier formula has been already known in 1881, Michelson was not able to make use of it. The more that he can not know then that the classical addition of velocities  $c+v$ is not valid in his device, not to mention his natural ignorance of the effect of Lorentz contraction of the longitudinal arm of the interferometer.

In the end of 1960-ies, when I performed measurements on modernized Michelson interferometers, the relativistic law $c/n\oplus v=c'$, where  $\oplus$ is the operator of relativistic addition of velocities and the condition $c'\leq c$ were already well known and described in all physics text-books. And so I used \cite{Demjanov second order} Fresnel formula for dependence of light's speed on parameters of a moving optical medium, by which the speed of light $c'$ in the medium never exceeds the speed of light $c$ in aether without particles:
\begin{equation}
c_\pm=\frac{c}{n}\pm v(1-\frac{1}{n^2}).\label{Fresnel}
\end{equation}
Following this line we will have instead of (\ref{longitudinal}) in the reference frame of stationary aether:
\begin{equation}
t_\|=\frac{l_\|}{c_+}+\frac{l_\|}{c_-}\label{longitudinal Fresnel}
\end{equation}
where $c_+$ and $c_-$ are values (\ref{Fresnel}) for the propagation of light in the light carrier along ${\bf v}$ and in the opposite direction respectively.

Detecting and measuring the shift of interference fringes is performed using tools situated in the movable Earth's frame of reference. Therefore passing in (\ref{longitudinal Fresnel}) to this reference frame it is necessary to make the correction on the effect of the Lorentz contraction of the length of the moving body. For the case in question we suppose that beyond the source of light and mirrors of the interferometer the light propagates mostly in the stationary aether perturbed by movable particles of a light carrier. It is here where all the time of delay  $t_\|$ of propagation of light in the arm $l_\|$ is  accumulated, that further is registered in the measurements of the interference fringe shift performed in the movable laboratory reference frame on the recording plate of the device.  And so in passing from aether to the Earth's laboratory reference frame of the experimental setup there should be taken into account the Lorentz contraction of the longitudinally movable arm $l$ of the interferometer:
\begin{equation}
l_\|=\frac{l}{\sqrt{1-v^2/c^2}}.\label{antiLorentz}
\end{equation}
Substituting (\ref{antiLorentz}) and respective values of (\ref{Fresnel}) into (\ref{longitudinal Fresnel}) we obtain for the direction parallel to ${\bf v}$
\begin{eqnarray}
t_{\|}&=&\frac{1}{\sqrt{1-v^2/c^2}}\left[\frac{l}{c_+}+\frac{l}{c_-}\right]=\frac{1}{\sqrt{1-v^2/c^2}}\frac{l}{c}\left[\frac{n}{1+\frac{v}{c}\frac{\Delta n^2}{n}}+\frac{n}{1-\frac{v}{c}\frac{\Delta n^2}{n}}\right]\nonumber\\
&\approx&(1+\frac{v^2}{2c^2})\frac{l}{c}\frac{2n}{1-\frac{v^2}{c^2}(\frac{\Delta n^2}{n})^2}\approx(1+\frac{v^2}{2c^2})\frac{l}{c}2n\left[{1+\frac{v^2}{c^2}\left(\frac{\Delta n^2}{n}\right)^2}\right]\nonumber\\
&\approx&\frac{l}{c}2n\left[1+\frac{v^2}{2c^2}+\frac{v^2}{c^2}\left(\frac{\Delta n^2}{n}\right)^2\right]\label{longitudinal Fresnel Lorentz}
\end{eqnarray}
where $\Delta n^2 = n^2-1$.

In the transverse arm the light propagates perpendicular to the direction of motion of the light carrier. In this case from (\ref{transverse}) and with the account of (\ref{Fresnel}) we obtain the time of delay $t_\perp$ of the ray. In the transverse to  $\mathbf{v}$ direction the length of the arm  ($l_\perp=l$) does not gain the Lorentz contraction, but formula (\ref{Fresnel}) takes into account the affect of movable particles of the optical medium (via the refractive index $n$ and value of the medium's permittivity  $\varepsilon=n^2$):
\begin{equation}
t_{\bot}=\frac{2l}{\sqrt{c^2/n^2-v^2}}=\frac{l}{c}\frac{2n}{\sqrt{1-\frac{v^2}{c^2}n^2}}\approx\frac{l}{c}2n\left(1+\frac{1}{2}\frac{v^2}{c^2}n^2\right).\label{transverse Fresnel}
\end{equation}
Subtracting (\ref{longitudinal Fresnel Lorentz}) from  (\ref{transverse Fresnel}) gives
\begin{equation}
\Delta t=t_\|-t_\perp\approx\frac{v^2}{c^2}\frac{l}{cn}\left[n^4-n^2-2(\Delta n^2)^2\right]=\frac{v^2}{c^2}\frac{l}{cn}\Delta n^2(n^2-2\Delta n^2)=\frac{v^2}{c^2}\frac{l}{cn}\Delta n^2(1-\Delta n^2).\label{Demjanov n}
\end{equation}
With $n=\sqrt{\varepsilon}$ and $\Delta\varepsilon=\Delta n^2$ we obtain from (\ref{Demjanov n}) the formula (Demjanov 1971) \cite{Demjanov}
\begin{equation}
\Delta t\approx\frac{v^2}{c^2}\frac{l}{c\sqrt{\varepsilon}}\Delta\varepsilon(1-\Delta\varepsilon).\label{Demjanov model}
\end{equation}

The difference $\Delta t$ obtained (\ref{Demjanov model}) corresponds to comparison of times $t_\|$ and $t_\perp$ of propagation of rays in a single arm of the interferometer in two orthogonal its positions, in one of which the arm is directed along $\mathbf{v}$, and in another orientation the same arm is directed perpendicular to $\mathbf{v}$. Really the construction of the interferometer has two orthogonal arms $l_1$ and $l_2$, whose rays interfere simultaneously at the screen of the device. The simultaneous occurrence on the interference screen of two orthogonal rays secures the continual observation of the very fringe and its shift ($\Delta X_m=ñX_o\Delta t/\lambda$) relative to itself. With this the formula for calculating the absolute shift of the fringe relative to itself at $90^\circ$ turning of the interferometer looks as follows:
\begin{equation}
\Delta X_m\approx X_0\frac{v^2}{c^2}\frac{l_1+l_2}{c\sqrt{\varepsilon}}\Delta\varepsilon(1-\Delta\varepsilon).\label{Demjanov model exp}
\end{equation}
Formula (\ref{Demjanov model exp}) is written down for "two-arms" case at $l_1 =l_2=l$, so it just corresponds to the experimental measurements of $\Delta X_m$. Formula (\ref{Demjanov model}) was deduced for a "single-arm" simplification of the reasoning, so it does not include in itself the coefficient 2.

Formula (\ref{Demjanov model exp}) reproduces in essential features the experimental curve (see Fig.\ref{fig1}). From this curve and theoretical model (\ref{Demjanov model exp}) there can be obtained the estimation of the velocity $v$ of  "aether wind". Curve 1 in Fig.\ref{fig1} corresponds to maximal values of the fringe shift, observed only about 1 hour  per day and night. The horizontal projection of the aether wind velocity is evaluated from this curve as $\sim480$ km/s. In the remaining time of day and night this projection is lesser that the maximum indicated. Twelve hours after observation of the maximal value, the horizontal projection of the velocity of aether wind passes the minimum $\sim140$  km/s. Thus, during 24-hour period of observation, the horizontal projection of the velocity of aether wind  at the latitude of Obninsk changes from 140 km/s to 480 km/s.

Accounting for that the term $1-n^{-2}=\Delta\varepsilon/\varepsilon$ determines the relative contribution of particles of the light carrying medium into its full permittivity $\varepsilon=1.+\Delta\varepsilon$ and that found by me effect of changing the sign of interference fringe shift is observed at $\Delta\varepsilon_\pm=1$ (see Fig.\ref{fig1}), it can be made up a notion about the contribution of two fundamental mechanisms of polarization of light carrying media. From Maxwell's formula for the full permittivity  ($\varepsilon =1.+\Delta\varepsilon$) of an optical medium the polarization contribution of aether $\varepsilon_{aether}=1.$, and the polarization contribution of particles $\Delta\varepsilon>0$. Since for $\Delta\varepsilon<1$ the sign $+\Delta X_m$ is positive, and for $\Delta\varepsilon>1$ the  sign $-\Delta X_m$ negative (Fig.\ref{fig1}), then the positive sign of the delay time $\Delta t=t_\perp-t_\|$ of rays by (\ref{Demjanov n}), because of the proportion $\Delta X_m\sim\Delta t$, when $\Delta\varepsilon<1$ is determined by the dominance of the polarization of aether in the full permittivity $\varepsilon =1.+\Delta\varepsilon$ of the light carrying medium, and when $\Delta\varepsilon>1$ $-$ by the dominance of the polarization of particles of light carrying medium.

\begin{figure}[h]
  \begin{center}
\includegraphics[scale=0.7]{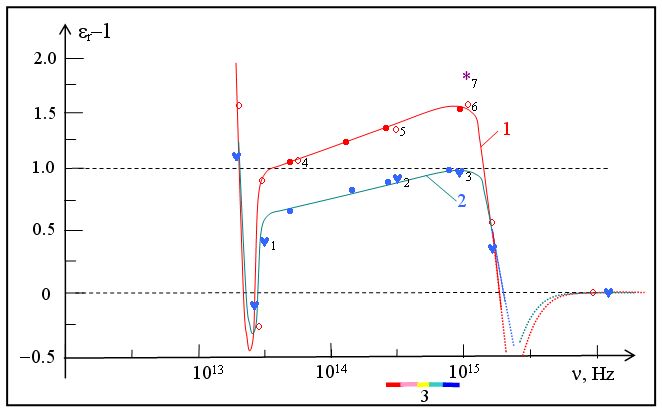}
  \caption{Frequency dependencies $\Delta\varepsilon(\nu)=\varepsilon_\textrm{r}(\nu)-1.$ of particles contribution $\Delta\varepsilon$ in dielectric permittivity ($\varepsilon_r=1+\Delta\varepsilon$) of fused quartz (curve 1) and distilled water (curve 2), measured in the range from  infra-red to ultra-violet; 3 is the range of visible part of the observation spectrum: hollow circles ($\circ$) and solid hearts ($\heartsuit$) are my measurements, a part of which up to the frequency $\sim3\cdot10^{13}$ Hz was published in the journal Izvestia Ac.Sci.USSR, ser. Inorganic materials, v.16, ¹5, 916 (1980); solid circles red ($\bullet$) and blue ($\bullet$) are data from \cite{Kaye} and \cite{Sosman}; points solid blue $\heartsuit_{1,2,3}$ and red $\circ_{4,5,6}$ were obtained on the Michelson interferometer used as a dielectrometer.
}\label{fig2}
\end{center}
\end{figure}

 Insofar as the information concerning frequency-dispersion properties of the materials studied is not well known I give in Fig.2 frequency dependencies of the dielectric permittivities of water and fused quartz in the region of normal dispersion squeezed  from both sides by the regions of abnormal dielectric dispersion. It presents known from literature and my experimental points of frequency-dispersion dependencies of the contribution $\Delta\varepsilon(\nu)=\varepsilon(\nu)-1.$ of particles in the fused quartz and water. The influence of losses from the regions of abnormal dielectric dispersion was overcome by such reduction of the length of light carriers (down to 10 cm, see table 1) when the possibility remained to register non-zeroth amplitudes $\Delta X_m$ of the harmonic shift of the interference fringe. Subtleties of the skill of of these observations are described in \cite{Demjanov experiment}.

\begin{table}
  \centering
\includegraphics[scale=0.6]{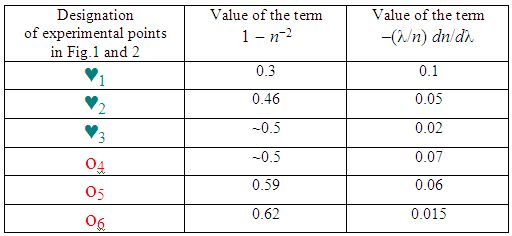}
  \caption{The comparison of contributions of the terms $1-n^{-2}$ and $-(\lambda/n)dn/d\lambda$
 of Fresnel formula for six experimental points of Fig.\ref{fig1}.}\label{table2}
\end{table}

In general case we must take into account the dispersion of the medium since the respective term, suggested by Lorentz,  enters the full Fresnel  formula $c'=c/n\pm v[1-1/n^2-(\lambda/n)dn/d\lambda]$. Values of the kinetic-polarization $(1-n^{-2})$ and kinetic-dispersion $(\lambda/n)dn/d\lambda$ terms of the Fresnel formula for six experimental points of Fig.\ref{fig1} are given in table 2. From table 2 wee see the leading role of the Fresnel term $(1-n^{-2})$ comparing with the kinetic-dispersion term $(\lambda/n)dn/d\lambda$. The accounting for the dispersion term in (\ref{Fresnel})-(\ref{Demjanov model}) gives the correction to the null point of $\Delta t(\Delta\varepsilon)$ as $\Delta\varepsilon=1+4(\lambda/n)dn/d\lambda$. Insofar as $dn/d\lambda<0$ this shifts, with the account of data from Table 1, the null point at 10-20$\%$ to $\Delta\varepsilon<1$. I has not been able to observe this effect  because the experimental errors were too large in order to make such a delicate observation.

After publication of formula (\ref{Demjanov model}), it has been deduced by P.C.Morris \cite{Morris} by another means $-$ from a generalization of Maxwell-Sellmeier formula onto moving optical media. This once again emphasizes the indispensable role of the corporeal medium (with the polarization contribution $\Delta\varepsilon$) in the formation of the law (\ref{Demjanov model}).

It should be noted that the authors \cite{Cahill} realized as well the necessity to perform the Michelson experiment in optical media. Though the linear model $\Delta X_m\sim l\Delta\varepsilon$ was obtained by them using the eclectic (non-Lorentz-invariant for effects of the second order by $v/c$) form $c/n\mp v$. Only using the invariant formula of the addition ($c/n\oplus v$) the light speed ($c/n$) in the optical medium with the velocity $v$ of particles of the medium relative to stationary aether we obtain the correct formula (\ref{Demjanov model exp}) for the shift of the interference fringe when turning the interferometer by $90^\circ$.

\section{Unnoticed detail capable to reduce and even nullify the shift $\Delta X_m$ of the interference fringe in the Michelson apparatus}

We see from the experiment that the sign of the shift of the interference fringe at $\Delta\varepsilon>1$ is opposite to that at $\Delta\varepsilon<1$. This means that being in the arms of the interferometer two kinds of the optical media will compete with each other. According to formula (\ref{Demjanov model exp}), which reproduces well this dependence, the effective contribution successively of the sections $l'$ and $l''$ with  respective $\Delta\varepsilon'<1$ and $\Delta\varepsilon''>1$ will be
\begin{equation}
l'\Delta\varepsilon'(1-\Delta\varepsilon')+l''\Delta\varepsilon''(1-\Delta\varepsilon'').\label{compete}
\end{equation}
Firstly I have confronted with this effect in 1970 when employed in the experiment with air arms the glass phase-changer. According to (\ref{compete}) at some interrelation of lengths, e.g. of the gas $l'$ and glass $l''$  sections, the phase shift (\ref{Demjanov model exp}) may be vanishing. Two important practical conclusions follow from this experimental observation.

1. The plates with $\Delta\varepsilon>1$ should not be used as phase-changers in gas-filled interferometers if the optical medium is a gas or water since for the glass $\Delta\varepsilon''(1-\Delta\varepsilon'')\approx-2$ while for the air $\Delta\varepsilon'(1-\Delta\varepsilon')\approx+0.0006$ and in visible range of light wavelength water has $\Delta\varepsilon'(1-\Delta\varepsilon')\approx+0.09\div+0.24$. So I used instead a water triplex plate comprised of the water layer of thickness $\sim 1$ mm located between two glass plates each of the thickness $\sim 0.1$ mm. The effective value $\Delta\varepsilon''(1-\Delta\varepsilon'')$ of such triplex is about $+0.15$, i.e. of the same sign as $\Delta\varepsilon'(1-\Delta\varepsilon')\approx+0.0006$ of the air.

2. In order to increase the effective length of the arms there frequently used configurations with zigzag path of light. Such construction demands the involving of many mirrors. Disposing this mirrors such as it is custom in ordinary life $-$ by the glass side facing to the incident light $-$ we come to the parasitic effect above mentioned: opposite signs of $\Delta\varepsilon(1-\Delta\varepsilon)$ in the air and glass compensate each other. As being evaluated by formula (\ref{compete}), that I firstly found experimentally \cite{Demjanov}, the thickness of glass layer $0.3$ mm cancels the phase difference produced by 1 m of the air path. Thus for the interferometer with $l=11$ m the six zigzag  reflecting mirrors of such thickness are capable to reduce three times the shift of interference fringe. For the thickness 0.5 mm of glass directed towards the incident light the shift of interference fringe will be nullified. In order to remove this harmful effect the glass mirrors should be oriented by the back side towards the light. In this event the amalgam layer should be thoroughly polished.

No wonder if there will come out dozens of similar interferometers in the world where the "zeroth" shift of interference fringe occurred because of the improper using the glass mirrors. Experiments performed on such interferometers contributed their portion of mischief into the myth about negative result of measuring the absolute motion. My experiments \cite{Demjanov}, \cite{Demjanov PLA} debunk these myths.

\section{Discussion of results and conclusion}

On the Michelson type interferometer there was obtained for optical media with the refractive index greater than one ($n>1$) the shift of the interference fringe that enhances considerably the ratio signal/noise. The positiveness (rather than negativeness) of Michelson experiments is thus demonstrated.  There was measured the difference  of round-trip times between the paths of propagation of light in parallel and transverse direction to translatory motion of the interferometer. It was found that this difference depends parabolically on the contribution $\Delta\varepsilon$ of the particles polarization into the full permittivity of the optical medium. Experimental observations are well described by the theoretical model based on the pre-relativistic (appeared to be Lorentz-invariant) and non-Galilean formula of Fresnel for the drag of light by the moving optical medium. Thus there was shown the inadequacy  of the additive rule $c+v$ for composition of the light's speed $c$ and the velocity $v$ of a moving inertial body used by  Michelson in deriving formula (\ref{Michelson}).

It should be stressed that in terms of the interpretation suggested, which is conceptually different from the standard one, all known measurements (see e.g. \cite{Michelson Morley, Miller}) on Michelson type interferometer, where the normal air pressure was maintained (that corresponds to $\Delta\varepsilon\approx 0.0006$), always would indicate $\Delta X_m\neq 0$ and hence $\Delta t\neq 0$. That is why all experimenters reported that in experiments with normal air pressure they noticed non-zeroth values $\Delta X_m$ among the noise. But Michelson missed $\Delta\varepsilon$ in his formula (\ref{Michelson}), and thus by the noticed trace of $\Delta X_m$ in air light carriers he obtained an  underestimation in $\Delta\varepsilon^{-1/2}\approx0.0006^{-1/2}=40$  times of  the absolute speed of the Earth. And so  all mentioned authors obtained instead of $200-400$ km/s estimations lying within the bounds of the noise, 5$-$10 km/s.

I took into account $\Delta\varepsilon$ in my formula (\ref{Demjanov model}). That saved from the fatal mistake, which omits the aether wind, and enabled me to detect the motion of the Earth with respect to stationary aether. The projection of the absolute velocity of the Earth on the horizontal plane of the experimental setup, calculated from (\ref{Demjanov model}), appeared to be in various times of day and night at the latitude of Obninsk  $140\div480$ km/s. Comparing (\ref{Michelson}) and (\ref{Demjanov model}) we see that accounting for $\Delta\varepsilon$ gives $1/\Delta\varepsilon_{air}\sim40$ times greater velocities of "aether wind" than Michelson$\&$Morley \cite{Michelson Morley},  Miller \cite{Miller} and others obtained from (\ref{Michelson}).

The watching of the discussion caused by the publication of my paper in PLA \cite{Demjanov PLA} shows that many (if not all) have forgotten that the first 40 years (after 1881) experiments of Michelson regarded negative because of that seeking to measure $\Delta X_m$ experimenters was not able to notice a weak harmonic shift $\Delta X_m$ of the fringe among the background noise \cite{Demjanov experiment}. In following 40 years (from 1920 to 1950), when owing to experiments of Miller \cite{Miller}  the fringe shift has been confidently observed, the fatal role in attribution the experiments of Michelson type to "negative" was played by 40 times understating error in Michelson formula (\ref{Michelson}) in comparison with the correct formula (\ref{Demjanov model}), rightly accounting for dielectric permittivity of the light carriers of the interferometer.

The disregard of the inertial contribution $\Delta\varepsilon$ by Michelson and all who copied him was the reason why famous  experiments on vacuumed interferometers and experiments with $\gamma$-rays gave negative results. Supposedly, before my experiments a medium (e.g. the air) was considered as an obstacle for the measurements, whose elimination allegedly would retain the shift of the interference fringe in "pure" form. This lead to what the expectations  by the formula (\ref{Michelson}) of the value of shift in vacuum were overstated about 1600 times comparing with that could be observed in the air \cite{Demjanov experiment}.  At the same time in vacuum (n=1), as I have shown for the first time experimentally \cite{Demjanov}, the shift is absent since $\Delta\varepsilon=0$. And since the corporeal medium is real environment is an equal participant in the process of forming the interference fringe shift, it is clear that the vacuumization of the light carrying parts of the interferometer vacuum did not confirm the results of the Michelson$\&$Morley \cite{Michelson Morley} and Miller \cite{Miller}, obtained in air of normal pressure.
 In all such experiments there was $\Delta\varepsilon=0$ either because of vacuum being in the part of interferometer where the light goes or due to vanishing $\Delta\varepsilon_\gamma$, since the dielectric permittivity $\varepsilon_\gamma$ of any optical media in $\gamma$-rays, i.e. $\Delta\varepsilon_\gamma=\varepsilon_\gamma-1\approx 1$. And I demonstrated by the direct experiments that when the air is evacuated from the regions where the light propagates the harmonic shift of the interference shift vanishes (become zero). Indeed, by (\ref{Demjanov model}) $\Delta\varepsilon\rightarrow 0$ entails  $\Delta t\rightarrow 0$, since  $\Delta X_m\sim\Delta t$.

The experimental results obtained by me and their theoretical interpretation bespeaks of the conventional character of the special relativity theory conditioned by the artificial extrapolation (arisen by chance in 1881 because of the ante-relativistic state of science) of the additive rule of the composition of velocities of inertial bodies to electromagnetic waves emitted by the inertial source. It is strange that formula (\ref{Michelson}) obtained by Michelson on the basis of the classical rule of the addition of velocities ($c'=c+v$) is not until now publicly acknowledged as a harsh error of the epoch 1881-1905 years. And yet the form $c'=c+v$ and formula (\ref{Michelson}) is not valid for the rotary interferometer with transverse light carrying arms also because they do not take into account the great sensibility of the measured by the interferometer parameter $\Delta t$ (i.e. the difference of times of propagation the longitudinal and transverse beams) to refractive index ($n>1$) of light carriers (and this is an irrefutable experimental fact). After all this device has not simply the dependence of $\Delta t$ (or $\Delta X_m$) on the refractive index $n$ of light carrying media of its arms, it exhibits the supersensibility of measurements $\Delta t$ (or $\Delta X_m$) to variations of the refractive index $var(n)$ of the optical media of each arm already in $4\div6$ signs after the point: $var_{xyz}(n)=1.000xyz$.

So that to go on regarding the experiments of Michelson type "negative in principle" in the period 1930-2010 years, when there have appeared convincing proofs of their "positiveness" in the sense of $\Delta X_m\ne0$ \cite{Michelson Morley, Miller}, \cite{Demjanov}-\cite{Demjanov first order} is the misfortune of the epoch. As a matter of fact, the special relativity appeared to be the artifact of this adversity. All the corpus of already published by me experimental results convincingly demonstrates  that  Michelson-type experiments are "positive" in principle since they register in Earth's laboratories $\Delta X_m\ne0$. Moreover, they are confidently reveals on the basis of that or another fixing $\Delta X_m\ne0$ the absolute motion of the Earth in the stationary aether with the value of horizontal projection of the velocity from 140 to 480 km/s registered in 24-hour cycle of observation in the day and night (at the latitude of Obninsk). After all in the orientation of the longitudinal arm of the interferometer to the Hercules constellation the Michelson type interferometer by any non-zeroth measurements $\Delta X_m\ne0$, processed with my formula (\ref{Demjanov model}), confidently state at any time of day and night and in any point of location the Earth's laboratory the velocity of the Earth relative to aether about 600 km/s \cite{Demjanov}.

\end{document}